\documentclass[aps,prl,twocolumn,floats,nofootinbib]{revtex4}
\usepackage{graphics,graphicx,epsfig}
\usepackage{amssymb,color}
\usepackage{epsf,epstopdf,wrapfig}
\usepackage {amsmath}

\newcommand{\beq}{\begin{equation}}
\newcommand{\eeq}{\end{equation}}
\newcommand{\beqn}{\begin{eqnarray}}
\newcommand{\eeqn}{\end{eqnarray}}

\begin{document}

\title{Optimal local estimates of visual motion in a natural environment}

\author{Shiva R.~Sinha,$^a$ William Bialek,$^b$ and Rob R.~de Ruyter van Steveninck$^a$}

\affiliation{$^a$Department of Physics, Indiana University, Bloomington IN 47405\\
$^b$Joseph Henry Laboratories of Physics, and Lewis--Sigler Institute for Integrative Genomics, Princeton University, Princeton NJ 08544\\
$^b$Initiative for the Theoretical Sciences, The Graduate Center, City University of New York, 365 Fifth Ave, New York NY 10016} 

\begin{abstract}
Many organisms, from flies to humans, use visual signals to estimate their motion through the world.  To explore the  motion estimation problem, we have constructed a camera/gyroscope system that allows us to sample, at high temporal resolution, the joint distribution of input images and rotational motions during a long walk in the woods.  From these data we construct the optimal estimator of velocity based on spatial and temporal derivatives of image intensity in small patches of the visual world.  Over the bulk of the naturally occurring dynamic range, the optimal estimator exhibits the same systematic errors seen in neural and behavioral responses, including the confounding of velocity and contrast.  These results suggest that apparent errors of sensory processing may reflect an optimal response to the physical signals in the environment.
\end{abstract}

\date{\today}

\maketitle

Humans and other animals use their visual systems to extract a wide variety of information about the world;
one such feature is motion.  A common theme in current models for visual motion computation is that these models do not, in general, produce a veridical estimate of the underlying velocities, and similar errors occur in neural and behavioral responses. In particular, the velocity of motion tends to be confounded with the spatial structure and even with the overall contrast of the visual scene. When the brain gets the wrong answer to a computational problem, it behooves us to ask why. One possibility is that this is a consequence of some fundamental limitation of the biological hardware---the correct computation simply is not realizable in cells and synapses that organize themselves during development. At the opposite extreme, it is possible that the brain performs a computation which is well matched to signals that it receives from the outside world, but that these physical data themselves typically are too limited to generate a reliable and correct answer \cite{bialek_12}. To locate the problem of visual motion estimation on this continuum from biological to physical limitations, we need to calibrate the data that the visual system uses as its input.

The modern discussion of models for visual motion estimation goes back to the work of Hassenstein and Reichardt in the 1950s, who studied the behavioral responses of beetles to simplified visual stimuli \cite{reichardt1,reichardt2}.  Early discussions of algorithms for motion estimation in mammalian visual system emphasized connections to the ``Reichardt correlator,'' although in mammalian visual cortex the rigid spatial sampling by the insect compound eye could be replaced by a more flexible set of receptive fields \cite{adelson+bergen_85,vansanten+sperling_85}.  In the blowfly visual system, single neurons encode motion estimates whose precision is close to the physical limits set by noise in the photoreceptor array \cite{rieke+al_97}, yet these same neurons exhibit the confounding of velocity and contrast predicted by the Reichardt model, at least at low contrast \cite{egelhaaf+al_89,ruyter+al_94}, and these same systematic errors are seen in the fly's behavior \cite{reichardt+poggio_76}.  In primate cortex, the variability of responses in single motion sensitive neurons makes a measurable contribution to the variability of perceptual judgements about motion direction \cite{britten+al_96}; again the responses of these neurons confound velocity and contrast, at least under some range of conditions \cite{heuer+britten_02}, as does human perception  \cite{vansanten+sperling_84}. While interest in understanding human vision has led to considerable focus on primates,  the insect visual system has returned as an important example in part because of genetic tools that make it possible to trace the complete circuits responsible for particular computations in the fruit fly \cite{rister+al_07,takemura+al_13,maisak+al_13,fisher+al_15}.  It seems an opportune time to ask not just what these circuits are computing, but why.

When data are noisy, the best estimate of an interesting feature is determined by the joint distribution of
that feature and the available input data. In the context of visual motion estimation, the feature is the
velocity of motion itself and the data are the `movies' collected by the eye. Our goal here is to sample this
joint distribution directly, and thus to construct empirically the function which optimally transforms visual in-
puts into motion estimates. We make progress on this seemingly daunting task by focusing on a small patch of
the visual world, and on situations where motion is dominated by rigid rotational of the observer, which can
be measured mechanically with a gyroscope.  In the first instance we are interested in problems of motion estimation in the 
fly visual system, although we discuss below the extent to which our data are relevant to primate and human vision. 

Photoreceptors in the fly's eye act as nearly perfect photon counters up to intensities corresponding to bright daylight, and
operate with a temporal bandwidth of $100 - 200\,{\rm Hz}$. The optics of the compound eye corresponds to a regular hexagonal lattice of receptors with horizontal spacing $\phi_0\sim 1.5^\circ$,  each of which looks out at the world through
a roughly Gaussian point spread function with a width $\sigma\sim 0.5^\circ$.  The class of photoreceptors that project into
the motion processing pathway have a spectral sensitivity that peaks at $\lambda_{\rm max} = 490\,{\rm nm}$.  To calibrate the data arriving at the fly's eye, we need to make a measurement that is better than what the fly does itself.

To exceed the signal-to-noise ratio of the fly's eye, we construct an imaging system with a single $25.4\,{\rm  mm}$ lens (Newport) which permits each device pixel to collect $\sim 7\times 10^5$ times as many photons as would one pixel in the fly's compound eye (with diameter of $30\, \mu{\rm m}$) under the same conditions (Fig 1A,B). The imaging process includes several components chosen to match the point spread function of our instrument to that of the  fly's receptors (Fig 1C). We use an array of photodetectors (Hamamatsu, S8729--10) with noise performance close to the photon shot noise limit when operated in daylight (Fig 1D), and we read out the photodetector signals at 1000 Hz \cite{calibrate}.  In the analysis described below, we smooth and downsample the data by a factor of two.

\begin{figure}
\includegraphics[width = \linewidth]{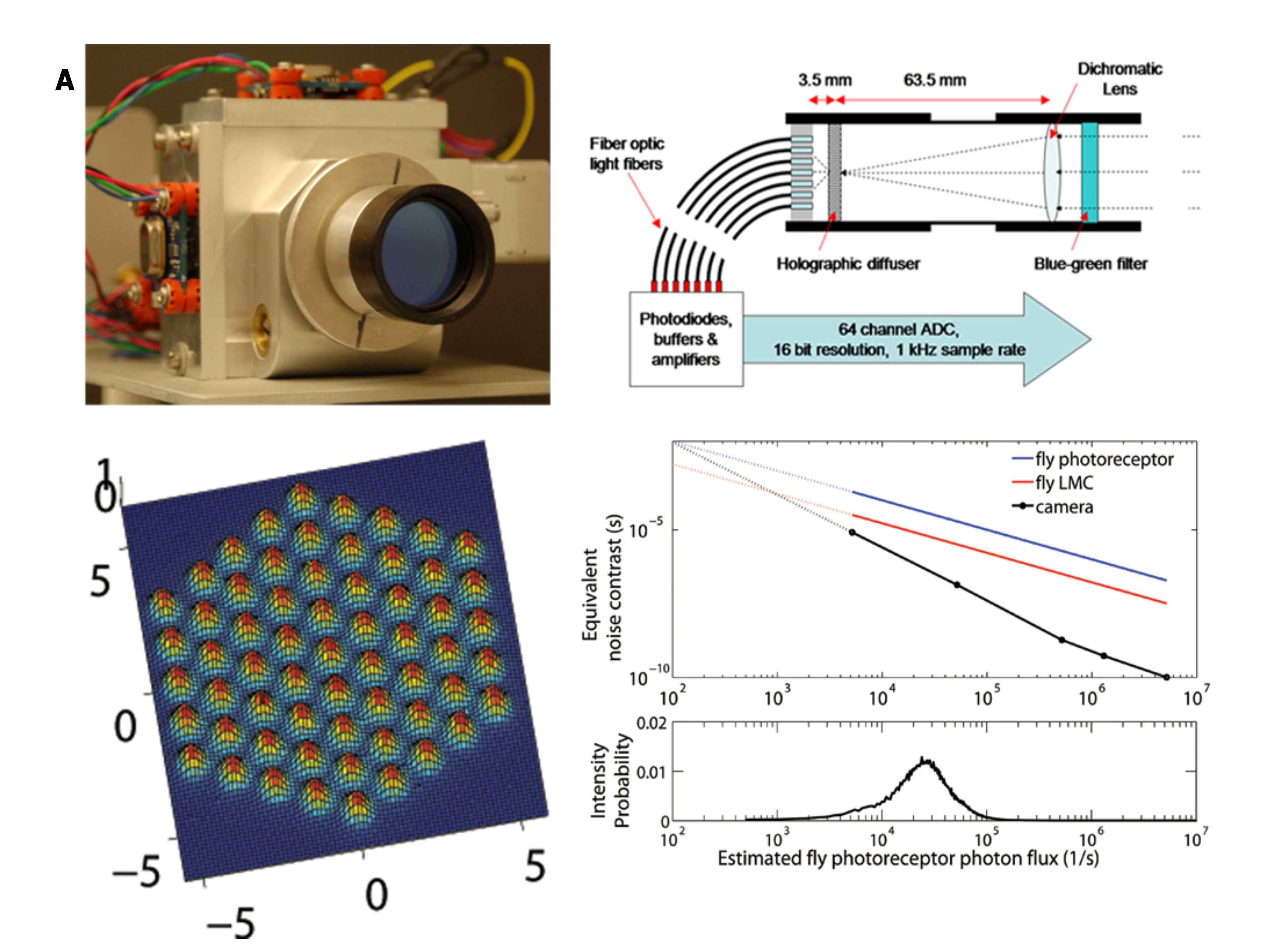}
\caption{Sampling movies  and   motion simultaneously. (A) Photograph of the camera headstage. Gyroscopes are mounted on orthogonal surfaces of the headstage, allowing the measurement of yaw, pitch, and roll angular velocities. The optical tube is centered in the headstage. (B) Schematic of the optical tube, in cross-section,  showing the focusing lens, spectral filter ($500\pm20\,{\rm nm}$), holographic diffuser, and hexagonal array of fiber optic cables (Edmund Optics, NT57--097). Not shown are the electronics for transducing the intensity and gyroscope signals and computer for recording the optical and gyroscope voltages. (C) Point spread functions of the 61 intensity channels and the relative position of channels in the hexagonal array. (D) Equivalent noise contrast of camera, demonstrating the much lower noise floor in the camera relative to the fly photoreceptors and large monopolar cells (LMC) under similar conditions.\label{camera}}
\end{figure}

For insects (and for us), motion has contributions from the animal's own movement and from the movements of objects in the environment.  On a relatively still day, with objects far away, self--motion is dominant.  Under these conditions we can measure angular motions directly with a gyroscope. We use three gyroscopes  (BEI Technology Inc, LCG50--00500--100 for yaw and pitch,  QRS130-01000--103 for roll) , aligned along the three cardinal axes, although our focus here is on the azimuthal or yaw motion. Since gyroscopes are sensitive to acceleration, some care is required to verify that we have the stability and noise performance required to integrate and generate a reliable velocity signal.  All data were recorded using a National Instruments PCI 6031 data acquisition card with 16--bit ADC resolution.

\begin{figure}[b]
\includegraphics[width = \linewidth]{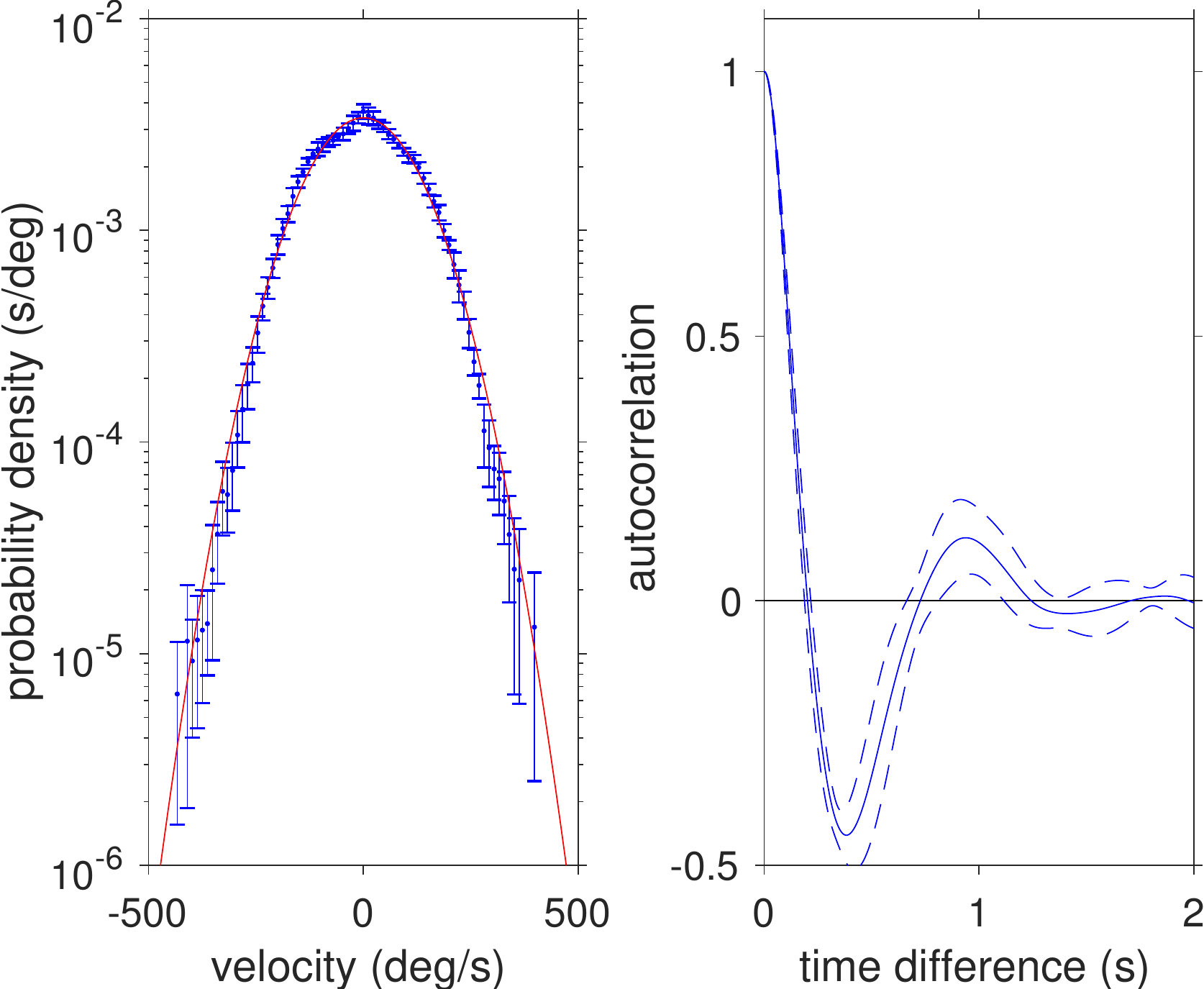}
\caption{Statistics of azimuthal (yaw) velocites.  (left) Probability distribution of instantaneous velocities.  Error bars are standard deviations across randomly chosen quarters of the half hour walk, and the red line is a Gaussian with the same mean and variance as the data. (right) Normalized autocorrelation function.  Dashed lines show $\pm$ one standard deviation across randomly chosen quarters of the data.  \label{vstats}}
\end{figure}

\begin{figure}[t]
\includegraphics[width = \linewidth]{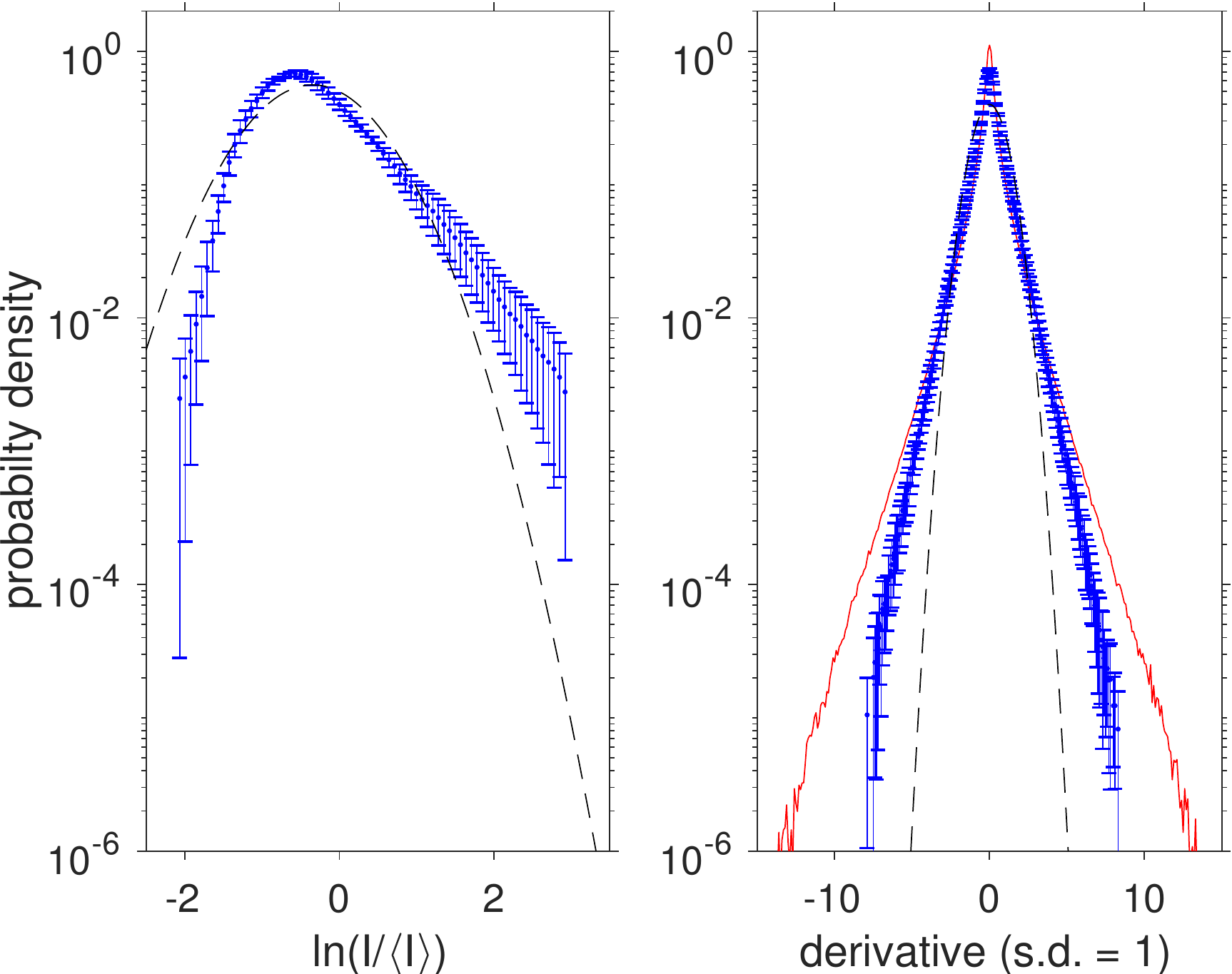}
\caption{Statistics of light intensities. (left) Distribution of (ln) intensity, collected over all 61 pixels.   (right) Distributions of spatial (blue) and temporal (red) derivatives of the (ln) intensity. Error bars are standard deviations across randomly chosen quarters of the half hour walk, and black dashed lines are Gaussians with the same mean and variance as the data.  \label{IIstats}}
\end{figure}

Ideally we would ``fly'' the instrument in Fig \ref{camera} along a trajectory taken by a real fly, but this is challenging.  As a first step, we take a half hour walk in the woods, letting the instrument hang from the arm of the person walking \cite{walk}.  Perhaps surprisingly, azimuthal  angular velocities along such a human--generated trajectory are quite large, with a standard deviation of $\sim 100^\circ/{\rm s}$ and a correlation time of $\sim 100\,{\rm ms}$ (Fig \ref{vstats}).  This scale of angular velocities is comparable to that of flies in reasonably straight flights, though not in acrobatic flight; the fluctuations are a bit slower and perhaps more Gaussian than  for flies.  The distribution of light intensities that we encounter on the walk is skewed, with roughly exponential tails, and this is even clearer in the distribution of (log) intensity gradients or time derivatives (Fig \ref{IIstats}); these basic features of natural scenes are familiar from earlier work  \cite{ruderman+bialek_94}.

The data we collect provides samples out of the joint distribution of movies and motions.  How do we relate these samples to the structure of the optimal motion estimator?  The fly visual system has access only to photoreceptor outputs, which we can write as the voltages $\{V_{\rm n}(t)\}$.  These are filtered, noisy versions of the light intensities in each pixel, $\{I_{\rm n}(t)\}$, which is what we measure with our camera.  These intensities in turn are related only probabilistically to the angular velocity $v(t)$; what we sample is the joint distribution $P[v(t), \{I_{\rm n}(t)\}]$.  All of the information about velocity is contained in the conditional distribution, $P[v(t)|\{V_{\rm n}(t)\}]$, and the structure of this distribution determines the computation that is needed in order to make optimal estimates \cite{potters+bialek_94}.    For some time it has been possible to characterize, quantitatively, the relationship between light intensity and voltage in the photoreceptors, but we have just had to guess at the joint distribution of intensities and velocities.  Our new data give us samples out of this distribution, and where the theory of optimal estimation asks for integrals over this distribution, we can approximate these integrals as sums over the measured samples, as in Monte Carlo simulations.

To focus more closely on what we learn from our new data, we can ignore the filtering and noise in the photoreceptors and imagine that the fly's brain has direct access to the light intensities $\{I_{\rm n}(t)\}$.   This is plausibly a good approximation under bright daylight conditions, and in a fuller analysis we can add back the receptor noise (much of which is photon shot noise) to see how the structure of the optimal estimator changes as light levels are lowered \cite{more}.  Notice that if we are searching for systematic errors of the motion computation that arise as the optimal response to physical limitations in the signal, then by using the actual intensity signals (rather than receptor voltages) we are being conservative.  With this approximation, the optimal estimate of velocity at a particular moment in time is 
\begin{equation}
\hat v_{\rm opt}(t_0) = \int dv\, P[v(t_0) = v|\{I_{\rm n}(t)\}] v  ,
\label{vhat1}
\end{equation}
where the notation reminds us that the optimal estimate depends on the pattern of light intensities over some window of time surrounding $t_0$.   This estimate is optimal in the sense that the mean--square error of our velocity estimates will be as small as possible, and the magnitude of these errors is determined by the width of the distribution $P[v(t_0) |\{I_{\rm n}(t)\}]$.

Equation (\ref{vhat1}) is complicated in part because the best estimate of velocity depends on the dynamics of light intensities in all the pixels.  In insects and in us, estimates of global rotational motion have long been thought to be built out of local motion estimates, and neurons that extract these local estimates now have been identified \cite{maisak+al_13}.  Because of the regular lattice structure of the insect visual system, ``local'' really refers to a single pixel and its near neighbors, and we expect that at high signal--to--noise ratio nothing would be gained by longer ranged comparisons \cite{potters+bialek_94}.  Out of these local neighborhoods we can build lattice approximations to the spatial derivatives in the two cardinal directions, and we can also compute temporal derivatives by comparing successive samples.    We expect that the information about motion in a particular direction will be dominated by the gradient in that direction; similarly we expect that estimates of velocity at one time are dominated by local time derivatives.  We also expect that velocity is not correlated with absolute intensity, so we look at derivatives of the log of the intensity.  Then the best local velocity estimate becomes
\begin{equation}
\hat v_{\rm opt} = \int dv\, P[v| \hat\partial_\phi \ln I(t) , \partial_t \ln I(t) ] v  ,
\label{vhat2}
\end{equation}
where $ \hat\partial_\phi$ is the lattice approximation to the spatial derivative in the same angular direction that we measure the velocity $v$.    Notice that this is a map from the plane $(\hat\partial_\phi \ln I(t) , \partial_t \ln I(t))$ to the velocity $\hat v_{\rm opt}$.

\begin{figure}
\includegraphics[width = \linewidth]{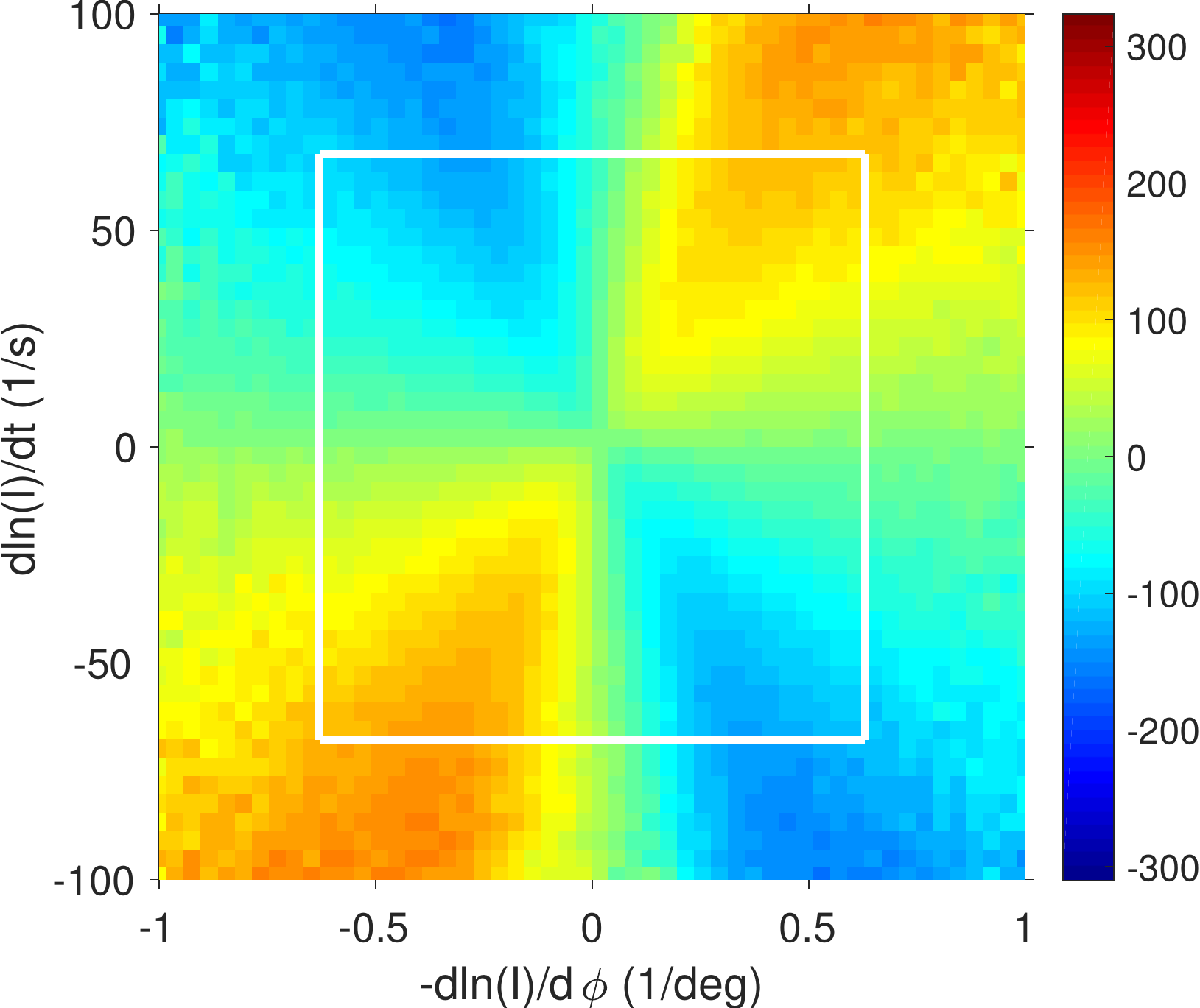}
\caption{Optimal estimator of velocity as a function of local spatial and temporal derivatives, from Eq (\ref{vhat2}). White box encloses 99\% of the data.  \label{vest99}}
\end{figure}

If we imagine a pattern moving rigidly across the array of detectors, then $I(\phi, t) = f(\phi - vt)$, and we can recover the velocity by taking a ratio of derivatives,
\begin{equation}
\hat v_{\rm grad} =- {{\partial_t \ln I(t) } \over {\hat\partial_\phi \ln I(t) } }  .
\label{vhat_grad}
\end{equation}
This ``gradient model'' of motion estimation gives veridical estimates in a mathematically idealized setting \cite{limb+murphy_75},  but the combination of differentiation and division makes it hugely susceptible to any random noise in the relation between measured intensities and velocities.  The Reichardt correlator estimates velocity as the product of neighboring pixel intensities, one of which is passed through a short time constant filter and one of which is passed through a longer time constant filter \cite{reichardt1,reichardt2,reichardt+poggio_76}; with proper (anti)symmetrization this approximates
\begin{equation}
\hat v_{\rm cor} \propto  {{\partial_t \ln I(t) } \times {\hat\partial_\phi \ln I(t) } }  .
\label{vhat_corr}
\end{equation}
With reasonable assumptions about the joint distribution of movies and motion, one can see both the gradient and correlation models as limiting cases of the general optimal estimator \cite{potters+bialek_94}, which suggests that known errors of motion estimation {\em could} be features of an optimal response to the available data, both in flies \cite{ruyter+al_94,bialek+ruyter_05} and  in humans \cite{weiss+al_02,stocker+simoncelli_06}.  But these theoretical approaches predict the observed errors of neural computation only in certain regimes, and to test the theory we need independent evidence that our visual systems really are operating in these limits.

The data from camera/gyroscope instrument can be thought of as giving us samples from the joint distribution $P[v, \hat\partial_\phi \ln I(t) , \partial_t \ln I(t)]$.  A practical version of Eq (\ref{vhat2}) is to discretize the measured gradients into bins, and then within each bin we compute the average velocity, resulting in map $\hat v(\hat\partial_\phi \ln I(t) , \partial_t \ln I(t))$ that is the optimal local motion estimator.  One caveat is that we can smooth the gradients in time to improve the quality of estimates, minimizing $\langle |\hat v - v|^2\rangle$ \cite{tau}.

The results for the optimal local motion estimator are shown in Fig \ref{vest99}.  We see that, at large values of the spatial gradient, contours of constant velocity are approximately linear, as expected in a gradient model [Eq (\ref{vhat_grad})].  But at smaller values of the spatial gradient, the contours of constant velocity bend into curves that approximate hyperbolae, which is what we expect in the correlator model [Eq (\ref{vhat_corr})].  Importantly, the bulk of the data that we collect on our half hour walk through the woods is in the regime where curvature of the constant velocity contours is prominent; we show this in Fig \ref{vest99} by outlining a region that encloses 99\% of the measured local gradients and time derivatives.

\begin{figure}[b]
\includegraphics[width = \linewidth]{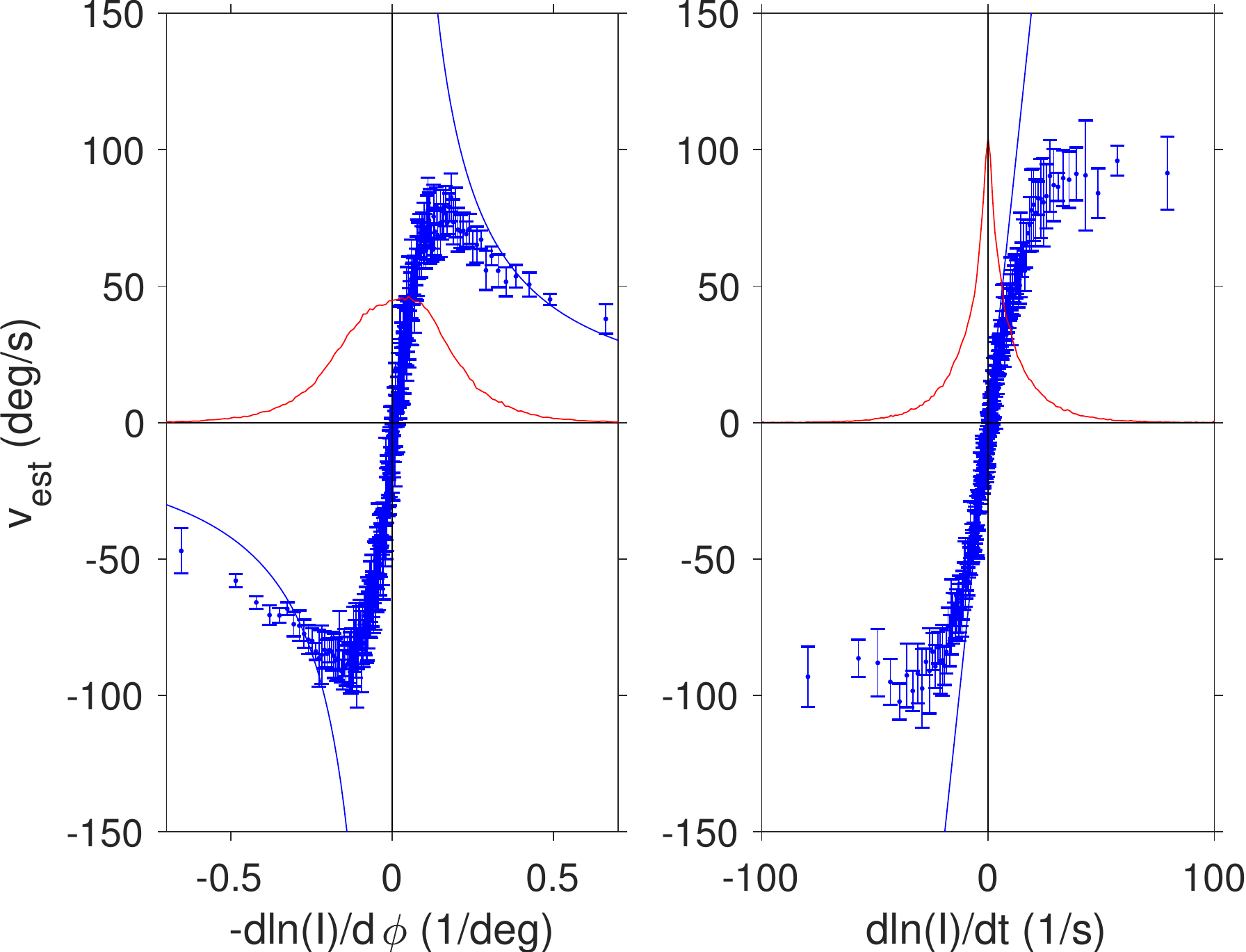}
\caption{Optimal motion estimates at constant temporal (left) or spatial (right) derivatives.  Blue points are opimal estimates, corresponding to slices through Fig \ref{vest99}; error bars are standard deviations across random quarters of the data.  Blue lines are the predictions of the gradient model, Eq (\ref{vhat_grad}), and red lines show the distribution of derivatives along the slice (scaled for clarity).  \label{slices}}
\end{figure}

We can get a clearer view of the optimal estimator by taking slices through Fig \ref{vest99}.  If we hold the time derivative of the local (log) light intensity fixed, and vary the spatial derviative, then correlator--like models predict that the velocity estimate will vary linearly [Eq (\ref{vhat_corr})], while gradient models predict that the estimate will vary inversely [Eq (\ref{vhat_grad})].  Both models predict that if we hold the spatial derivative fixed, the estimate should vary linearly with the temporal derivative.  In Fig \ref{slices} we see linear dependences at small values of the derivatives, along both slices.  With the temporal derivative held fixed, we see signs of the inverse dependence on the spatial derivative expected in the gradient model, but only at large derivatives, in the tail of the distribution that we encountered along our walk in the woods.

Optimal estimation quite generally is a tradeoff between systematic and random errors.  Thus, systematic errors are optimal only if they protect the estimate from random errors, and if the systematic errors are large this must be in response to large sources of randomness in the raw data on which estimates are based.  In our data, noise in the measurement of light intensity and its derivatives is small, by construction, so any ``noise'' is inherent in the probabilistic relationship between movies and motion.  If this effective noise is large enough to drive the optimal estimator into the correlator regime, then even the optimal estimates themselves should be noisy.  

\begin{figure}
\centerline{\includegraphics[width = 0.9 \linewidth]{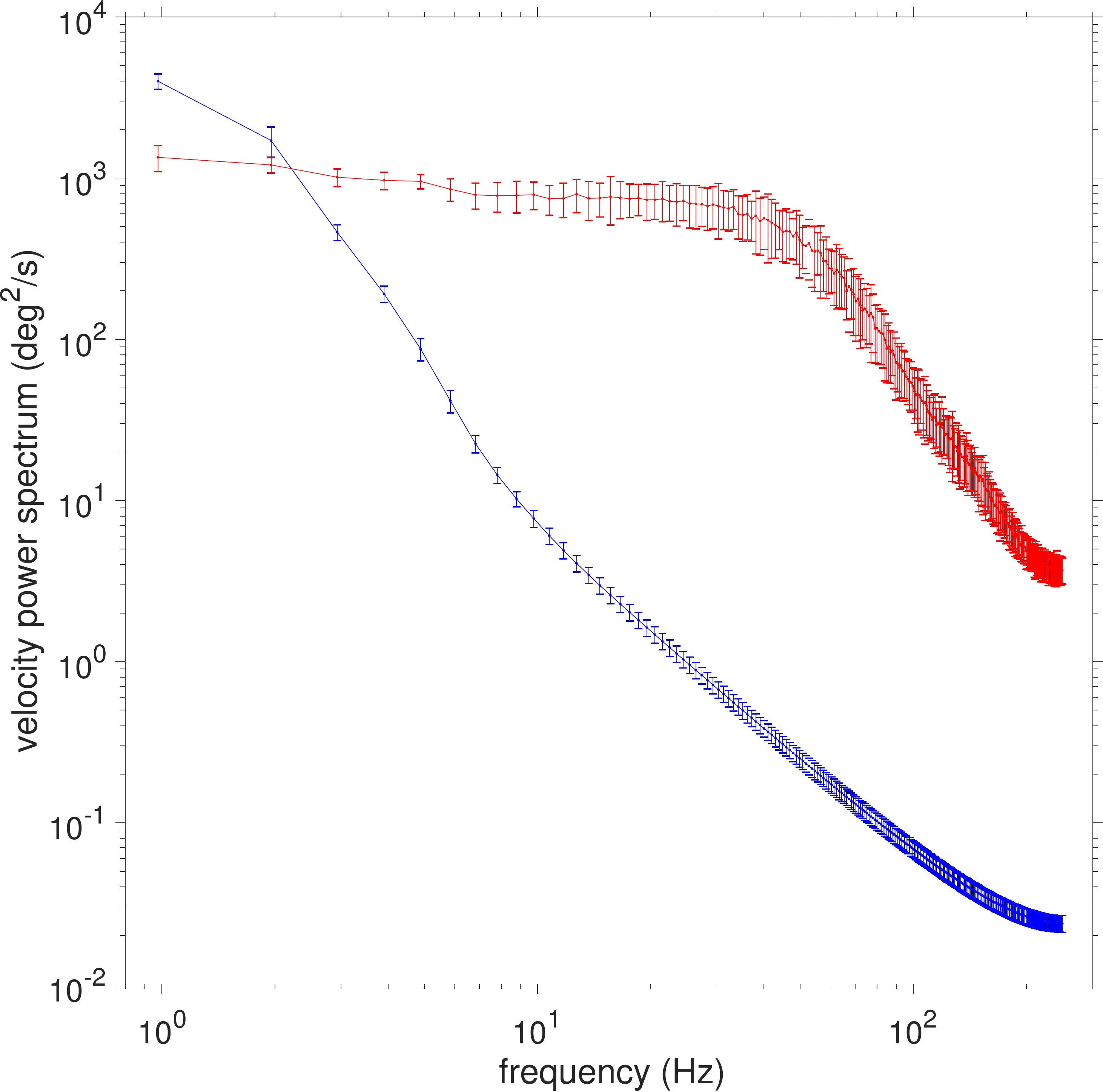}}
\caption{Power spectra of the velocity signal (blue) and  the effective noise in the optimal local estimator (red).  Note that this is in the absence of added photoreceptor noise.  Error bars are standard deviations across random quarters of the data.  \label{spectra}}
\end{figure}

To measure the noise in the optimal estimates, we use the estimator in Fig \ref{vest99}, together with the observed time series of spatial and temporal gradients, to generate a time series of velocity estimates $v_{\rm est}(t) \equiv \hat v$ that can be compared with the actual velocity $v(t)$.  In small time windows ($1.024\,{\rm s}$) we can compare the Fourier components of these data, and  find that the relationship is approximately linear,
\begin{equation}
\tilde v_{\rm est} (\omega) = g(\omega) \left [ \tilde v (\omega)  + \tilde\eta (\omega) \right].
\end{equation}
This definition of an effective noise $\tilde\eta (\omega) $ follows the usual strategy of referring noise to the ``input'' of the system, so that it can be compared meaningfully with the signal \cite{bialek_12}.  The (properly normalized) variance of $\tilde v (\omega) $ gives the power spectrum of the velocity signal, and the variance of $\tilde\eta (\omega) $ gives the power spectrum of noise in our estimates; results are shown in Fig \ref{spectra}.    Consistent with the correlator--like structure of the optimal estimator, signal--to--noise ratios are low, rising above unity only below $2\,{\rm Hz}$, to a maximum of $\sim 3$.

To summarize, the relationship between the local dynamics of images and movement velocities is sufficiently probabilistic that optimal estimates are driven into a regime where systematic errors are significant.  In this regime, the optimal estimator is approximately a correlator or motion energy estimator.    We have emphasized the connection to fly vision, but primate vision must make use of the same physical signals, and so the same considerations apply when the visual cortex is computing motion on the scale of $\sim 1.5 -3^\circ$ \cite{britten+heuer_99}.  The idea that apparent errors of motion computation might be optimal responses to physically limited signals is an old one, both in flies \cite{potters+bialek_94} or in humans \cite{weiss+al_02,stocker+simoncelli_06}, but as far we know Fig \ref{vest99} provides the first direct evidence that motion estimation in a naturalistic context really is in the regime where correlation is optimal.

The theory of optimal estimation involves integrals over the relevant probability distributions of raw data and interesting features.  Emphasizing the connections between statistical physics and statistical inference \cite{grassberger+nadal_94}, our approach replaces these integrals with sums over samples, as in Monte Carlo simulations, but in this case samples are drawn from the natural environment.   Along this path, we expect that there is much more to be done.  The crossover between correlator--like and gradient--like estimation should depend on the overall signal--to--noise ratio, which we can vary by adding back photon shot noise or focusing on periods with different typical values of image contrast.  Asymmetries in the underlying distributions should lead to asymmetries in the optimal estimator \cite{fitzgerald+al_11}, which are barely visible in Fig \ref{vest99} and should be connected to the separate processing of on and off signals \cite{behnia+al_14}.  It also will be interesting to understand the rules for optimal combination of these local estimators into wide--field motion signals.  Having seen that correlator--like motion computations can be optimal under natural conditions, the most important challenge is to make a more detailed comparison between the structure of these estimators and the structure of neural computation under conditions where the visual system is adapted to the same stimulus ensemble.

\begin{acknowledgments}
This work was supported in part by the National Science Foundation, through the Center for the Physics of Biological Function (PHY--1734030) and grants IIS--0423039 and PHY--1607612, and by the WM Keck Foundation.
\end{acknowledgments}


\begin{thebibliography}{99}
\bibitem{calibrate}
To spatially calibrate the optical channels, a moving dot was scanned in a raster pattern on a Tektronix 608 oscilloscope. The camera optics were aimed at the oscilloscope and the voltages recorded. Data similar to Fig \ref{camera}C were generated based on this calibration. The temporal response and gain characteristics were calibrated by generating a white noise sequence on a high intensity LED, and recording from optical channels.  Based on this characteristic an equivalent photon count rate was estimated.
\bibitem{walk}
SRS took the walk, in Dunn's Woods, a wooded area on the campus of Indiana University Bloomington.  We chose a sunny, cloud free summer day, around noon, for the brightest conditions possible.
\bibitem{tau}
For the data shown here, this optimal smoothing time is $\tau \sim 10\,{\rm ms}$, which is close to the shortest time constants observed for the transients responses of  motion sensitive neurons in the blowfly \cite{ruyter+al_94,ruyter+al_86}.  The optimal value of $\tau$ gets longer if we add back the effects of photon shot noise.
\end{thebibliography}
\end{document}